\documentclass[superscriptaddress,prl,amsmath,amssymb,bibnotes,altaffilletter,twocolumn]{revtex4-1}

\usepackage{lineno}
\usepackage{longtable}  
\usepackage{graphicx}
\usepackage{amsmath,amssymb,bbold,bm}
\usepackage{epstopdf}
\usepackage{xcolor}
\usepackage{lipsum}
\usepackage{slashed}

\begin{document}

\def\nuc#1#2{${}^{#1}$#2}
\def\mee{$\langle m_{\beta\beta} \rangle$}
\def\mnu{$m_{\nu}$}
\def\ml{$m_{lightest}$}
\def\gnu{$\langle g_{\nu,\chi}\rangle$}
\def\mmod{$\| \langle m_{\beta\beta} \rangle \|$}
\def\mb{$\langle m_{\beta} \rangle$}
\def\BBz{$\beta\beta(0\nu)$}
\def\BBm{$\beta\beta(0\nu,\chi)$}
\def\BBt{$\beta\beta(2\nu)$}
\def\nonubb{$\beta\beta(0\nu)$}
\def\twonubb{$\beta\beta(2\nu)$}
\def\BB{$\beta\beta$}
\def\Mz{$M_{0\nu}$}
\def\Mt{$M_{2\nu}$}
\def\MzG{$M^{GT}_{0\nu}$}           
\def\MzF{$M^{F}_{0\nu}$}                
\def\MtG{$M^{GT}_{2\nu}$}           
\def\MtF{$M^{F}_{2\nu}$}                
\def\Gz{$G_{0\nu}$}					
\def\Tz{$T^{0\nu}_{1/2}$}
\def\Tt{$T^{2\nu}_{1/2}$}
\def\Tc{$T^{0\nu\,\chi}_{1/2}$}
\def\Rz{$\Gamma_{0\nu}$}            
\def\Rt{$\Gamma_{2\nu}$}            
\def\ms{$\delta m_{\rm sol}^{2}$}
\def\ma{$\delta m_{\rm atm}^{2}$}
\def\mot{$\delta m_{12}^{2}$}
\def\mtt{$\delta m_{23}^{2}$}
\def\ts{$\theta_{\rm sol}$}
\def\ta{$\theta_{\rm atm}$}
\def\ttwo{$\theta_{12}$}
\def\tot{$\theta_{13}$}
\def\gpp{$g_{pp}$}                  
\def\gA{$g_{A}$}                  
\def\qval{$Q_{\beta\beta}$}                 
\def\be{\begin{equation}}
\def\ee{\end{equation}}
\def\cpKkgy{cnts/(keV kg y)}
\def\cpKkgd{cnts/(keV kg d)}
\def\cpRty{cnts/(ROI t y)}
\def\onecpRty{1~cnt/(ROI t y)}
\def\threecpRty{3~cnts/(ROI t y)}
\def\ppc{P-PC}                          
\def\nsc{N-SC}                          
\def\cosixty{$^{60}Co$}
\def\thttt{$^{232}\mathrm{Th}$}
\def\utte{$^{238}\mathrm{U}$}
\def\mubqkg{$\mu\mathrm{Bq/kg}$}
\def\cusulfate{$\mathrm{CuSO}_4$}
\def\MJ{{\sc Majorana}}             
\def\DEM{{\sc Demonstrator}}             
\def\MJDEMbf{\bfseries{\scshape{Majorana Demonstrator}}}
\def\MJbf{\bfseries{\scshape{Majorana}}}
\def\MJDEMit{\itshape{\scshape{Majorana Demonstrator}}}
\newcommand{\Gerda}{GERDA}
\newcommand{\GF}{\textsc{Geant4}}
\newcommand{\MaGe}{\textsc{MaGe}}

\title{Search for Tri-Nucleon Decay in the \MJ\ \DEM}

\newcommand{\blhill}{Department of Physics, Black Hills State University, Spearfish, SD, USA}
\newcommand{\ITEP}{National Research Center ``Kurchatov Institute'' Institute for Theoretical and Experimental Physics, Moscow, Russia}
\newcommand{\JINR}{Joint Institute for Nuclear Research, Dubna, Russia}
\newcommand{\lbnl}{Nuclear Science Division, Lawrence Berkeley National Laboratory, Berkeley, CA, USA}
\newcommand{\lanl}{Los Alamos National Laboratory, Los Alamos, NM, USA}
\newcommand{\queens}{Department of Physics, Engineering Physics and Astronomy, Queen's University, Kingston, ON, Canada} 
\newcommand{\uw}{Center for Experimental Nuclear Physics and Astrophysics, 
and Department of Physics, University of Washington, Seattle, WA, USA}
\newcommand{\unc}{Department of Physics and Astronomy, University of North Carolina, Chapel Hill, NC, USA}
\newcommand{\duke}{Department of Physics, Duke University, Durham, NC, USA}
\newcommand{\ncsu}{Department of Physics, North Carolina State University, Raleigh, NC, USA}	
\newcommand{\ornl}{Oak Ridge National Laboratory, Oak Ridge, TN, USA}
\newcommand{\ou}{Research Center for Nuclear Physics, Osaka University, Ibaraki, Osaka, Japan}
\newcommand{\pnnl}{Pacific Northwest National Laboratory, Richland, WA, USA}
\newcommand{\princeton}{Department of Physics, Princeton University, Princeton, NJ, USA}
\newcommand{\ttu}{Tennessee Tech University, Cookeville, TN, USA}
\newcommand{\sdsmt}{South Dakota School of Mines and Technology, Rapid City, SD, USA}
\newcommand{\usc}{Department of Physics and Astronomy, University of South Carolina, Columbia, SC, USA}
\newcommand{\usd}{Department of Physics, University of South Dakota, Vermillion, SD, USA} 
\newcommand{\ut}{Department of Physics and Astronomy, University of Tennessee, Knoxville, TN, USA}
\newcommand{\tunl}{Triangle Universities Nuclear Laboratory, Durham, NC, USA}
\newcommand{\mpi}{Max-Planck-Institut f\"{u}r Physik, M\"{u}nchen, Germany}
\newcommand{\tum}{Physik Department, Technische Universit\"{a}t, M\"{u}nchen, Germany}
\newcommand{\MIT}{Department of Physics, Massachusetts Institute of Technology, Cambridge, MA, USA} 

\affiliation{\uw}
\affiliation{\pnnl}
\affiliation{\usc}
\affiliation{\ornl}
\affiliation{\ITEP}
\affiliation{\usd}
\affiliation{\queens} 
\affiliation{\sdsmt}
\affiliation{\mpi}
\affiliation{\JINR}
\affiliation{\duke}
\affiliation{\tunl}
\affiliation{\uw}
\affiliation{\unc}
\affiliation{\lbnl}
\affiliation{\lanl}
\affiliation{\ut}
\affiliation{\ou}
\affiliation{\princeton}
\affiliation{\ncsu}
\affiliation{\MIT}
\affiliation{\blhill}
\affiliation{\ttu}
\affiliation{\tum}

\author{S.I.~Alvis}\affiliation{\uw}	
\author{I.J.~Arnquist}\affiliation{\pnnl} 
\author{F.T.~Avignone~III}\affiliation{\usc}\affiliation{\ornl}
\author{A.S.~Barabash}\affiliation{\ITEP}
\author{C.J.~Barton}\affiliation{\usd}	
\author{V.~Basu}\affiliation{\queens} 
\author{F.E.~Bertrand}\affiliation{\ornl}
\author{B.~Bos}\affiliation{\sdsmt} 
\author{V.~Brudanin}\affiliation{\JINR}
\author{M.~Busch}\affiliation{\duke}\affiliation{\tunl}	
\author{M.~Buuck}\affiliation{\uw}  
\author{T.S.~Caldwell}\affiliation{\unc}\affiliation{\tunl}	
\author{Y-D.~Chan}\affiliation{\lbnl}
\author{C.D.~Christofferson}\affiliation{\sdsmt} 
\author{P.-H.~Chu}\affiliation{\lanl} 
\author{C. Cuesta}\altaffiliation{Present address: Centro de Investigaciones Energ\'{e}ticas, Medioambientales y Tecnol\'{o}gicas, CIEMAT 28040, Madrid, Spain}\affiliation{\uw}
\author{J.A.~Detwiler}\affiliation{\uw}	
\author{Yu.~Efremenko}\affiliation{\ut}\affiliation{\ornl}
\author{H.~Ejiri}\affiliation{\ou}
\author{S.R.~Elliott}\affiliation{\lanl}
\author{T.~Gilliss}\affiliation{\unc}\affiliation{\tunl}  
\author{G.K.~Giovanetti}\affiliation{\princeton}  
\author{M.P.~Green}\affiliation{\ncsu}\affiliation{\tunl}\affiliation{\ornl}   
\author{J.~Gruszko}\affiliation{\MIT} 
\author{I.S.~Guinn}\affiliation{\uw}		
\author{V.E.~Guiseppe}\affiliation{\usc}	
\author{C.R.~Haufe}\affiliation{\unc}\affiliation{\tunl}	
\author{R.J.~Hegedus}\affiliation{\unc}\affiliation{\tunl} 
\author{L.~Hehn}\affiliation{\lbnl}	
\author{R.~Henning}\affiliation{\unc}\affiliation{\tunl}
\author{D.~Hervas~Aguilar}\affiliation{\unc}\affiliation{\tunl} 
\author{E.W.~Hoppe}\affiliation{\pnnl}
\author{M.A.~Howe}\affiliation{\unc}\affiliation{\tunl}
\author{K.J.~Keeter}\affiliation{\blhill}
\author{M.F.~Kidd}\affiliation{\ttu}	
\author{S.I.~Konovalov}\affiliation{\ITEP}
\author{R.T.~Kouzes}\affiliation{\pnnl}
\author{A.M.~Lopez}\affiliation{\ut}	
\author{R.D.~Martin}\affiliation{\queens}	
\author{R.~Massarczyk}\affiliation{\lanl}		
\author{S.J.~Meijer}\affiliation{\unc}\affiliation{\tunl}	
\author{S.~Mertens}\affiliation{\mpi}\affiliation{\tum}		
\author{J.~Myslik}\affiliation{\lbnl}		
\author{G.~Othman}\affiliation{\unc}\affiliation{\tunl} 
\author{W.~Pettus}\affiliation{\uw}	
\author{A.~Piliounis}\affiliation{\queens} 
\author{A.W.P.~Poon}\affiliation{\lbnl}
\author{D.C.~Radford}\affiliation{\ornl}
\author{J.~Rager}\affiliation{\unc}\affiliation{\tunl}	
\author{A.L.~Reine}\affiliation{\unc}\affiliation{\tunl}	
\author{K.~Rielage}\affiliation{\lanl}
\author{N.W.~Ruof}\affiliation{\uw}	
\author{B.~Shanks}\affiliation{\ornl}	
\author{M.~Shirchenko}\affiliation{\JINR}
\author{D.~Tedeschi}\affiliation{\usc}		
\author{R.L.~Varner}\affiliation{\ornl}  
\author{S.~Vasilyev}\affiliation{\JINR}	
\author{B.R.~White}\affiliation{\lanl}	
\author{J.F.~Wilkerson}\affiliation{\unc}\affiliation{\tunl}\affiliation{\ornl}    
\author{C.~Wiseman}\affiliation{\uw}		
\author{W.~Xu}\affiliation{\usd} 
\author{E.~Yakushev}\affiliation{\JINR}
\author{C.-H.~Yu}\affiliation{\ornl}
\author{V.~Yumatov}\affiliation{\ITEP}
\author{I.~Zhitnikov}\affiliation{\JINR} 
\author{B.X.~Zhu}\affiliation{\lanl} 
						
\collaboration{{\sc{Majorana}} Collaboration}
\noaffiliation


\begin{abstract}
The \MJ\ \DEM\ is an ultra low-background experiment searching for neutrinoless double-beta decay in $^{76}$Ge. The heavily shielded array of germanium detectors, placed nearly a mile underground at the Sanford Underground Research Facility in Lead, South Dakota, also allows searches for new exotic physics. We present the first limits for tri-nucleon decay-specific modes and invisible decay modes for Ge isotopes. We find a half-life limit of $4.9\times10^{25}$ yr for the decay \nuc{76}{Ge}(ppn) $\rightarrow$ \nuc{73}{Zn} $ e^+\pi^+$ and $4.7\times10^{25}$ yr for the decay \nuc{76}{Ge}(ppp)$\rightarrow$ \nuc{73}{Cu} $ e^+\pi^+\pi^+$. The half-life limit for the invisible tri-proton decay mode of \nuc{76}{Ge} was found to be $7.5\times10^{24}$ yr.

\end{abstract}


\date{\today}
\maketitle
\section{Introduction}
The conservation of the number of baryons ($B$) in any reaction is an empirical symmetry of the Standard Model that is not the result of any fundamental principle. Hence, there are numerous reasons to consider its violation ($\slashed{B}$). Theories that unify the strong and electroweak forces naturally include $\slashed{B}$. It is expected that quantum gravity theories will violate $B$ or any similar global symmetry. Theories with extra dimensions permit particle disappearance, and nucleon decay can be induced via interactions with dark matter as manifest in asymmetric dark matter theories. $\slashed{B}$ is also one of the Sakharov requirements to explain the matter-antimatter asymmetry of the Universe. These topics and the possibility of $\slashed{B}$ are reviewed in Ref.~\cite{Babu2013} and references therein. Therefore, the scientific motivation for studying $\slashed{B}$ is compelling. The breadth of model possibilities is very broad, however, indicating that many complementary search techniques could help elucidate the question. 

The Standard Model with small neutrino masses has an anomaly-free $Z_6$ symmetry that acts as discrete $B$~\cite{Babu2003}. In this model $\Delta B$=1 or 2 processes are forbidden, but $\Delta B$=3 transitions can arise due to a dimension 15 operator.  When undergoing a $\Delta B$=3 tri-nucleon decay, three baryons disappear from the nucleus, frequently leaving an isotope that is unstable. Previous searches in Xe isotopes~\cite{Bernabei2006,Albert2018} and \nuc{127}{I}~\cite{Hazama1994} looked for {\em invisible} decay channels assuming no observation of the initial tri-nucleon decay or disappearance.  Only the decay of the unstable product was sought as evidence for the process. Other groups considered invisible $\Delta B$=2 decays with limits reported in Refs.~\cite{Berger1991,Bernabei2000,Back2003,Tretyak2004,Araki2006,Litos2014,Takhistov2015,Gustafson2015}.  Results for $\Delta B$=2, 3 decays from the \MJ\ \DEM\ are presented here for invisible channels and for decay-specific modes.

\begin{figure}[h]
 \centering
 \includegraphics[width=0.75\columnwidth, keepaspectratio=true]{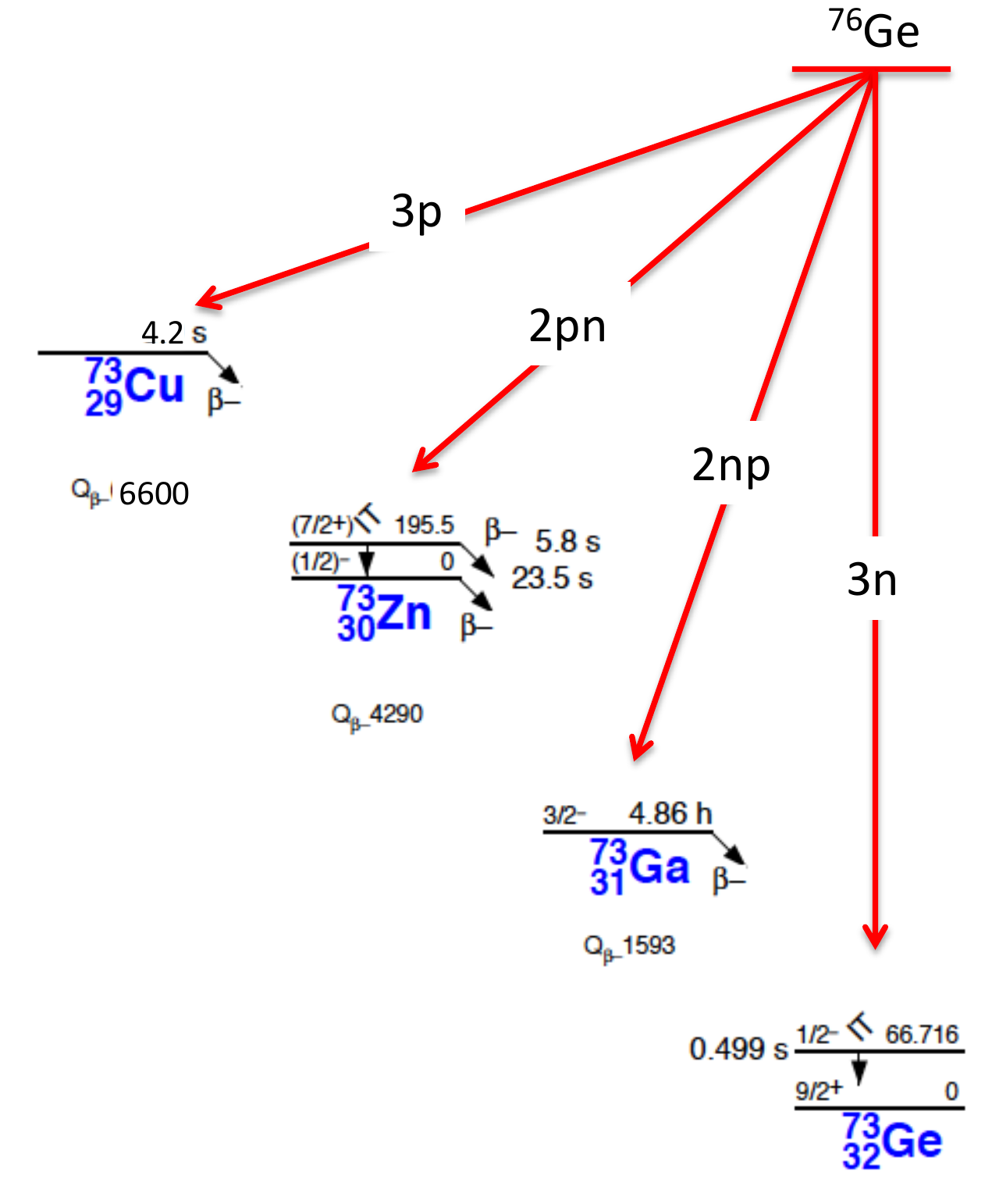}
 \caption{ The \nuc{76}{Ge} decay scheme. Figure adapted from Ref.~\cite{Firestone1997}. }
 \label{Fig:DecayScheme}
\end{figure}

The dominant decay modes for $\Delta B$=3 are given in Ref.~\cite{Babu2003} as
\begin{eqnarray}
\label{eqn:ThreeParticle}
ppp & \rightarrow  & e^+ \pi^+ \pi^+  \\ \nonumber
ppn & \rightarrow & e^+ \pi^+ \\ \nonumber
pnn & \rightarrow & e^+ \pi^0  \\ \nonumber
nnn & \rightarrow & \bar{\nu} \pi^0.
\end{eqnarray}
The resulting daughter nuclei for these processes in \nuc{76}{Ge} are displayed in Fig.~\ref{Fig:DecayScheme}. Typical modes of decay for $\Delta B$=2 are
\begin{eqnarray}
\label{eqn:TwoParticle}
pp & \rightarrow  &  \pi^+ \pi^+  \\ \nonumber
pn & \rightarrow & \pi^0 \pi^+ \\ \nonumber
nn & \rightarrow & \pi^+\pi^-, \pi^0\pi^0.
\end{eqnarray}

\section{The \textsc{Majorana Demonstrator}}
The \MJ\ \DEM\ described in detail in Refs.~\cite{Abgrall2014, Aalseth2018} is located at a depth of 4850\,ft at the Sanford Underground
Research Facility in Lead, South Dakota \cite{Heise2015}. In addition to its primary goal of searching for neutrinoless double-beta decay, its ultra low-background configuration permits additional physics studies including  searches for dark matter~\cite{Abgrall2017}, axions, and exotic physics (e.g. Ref.~\cite{Alvis2018}). Two modules contain 44.1\,kg of high-purity germanium P-type point-contact detectors, of which 29.7\,kg have 88\% $^{76}$Ge enrichment. Fifty-eight detector units are installed in strings of three, four, or five detectors.  These strings of detectors are mounted within vacuum cryostats which are shielded from room background by a lead and copper shield. The entire apparatus is contained within a 4-$\pi$ cosmic ray veto system~\cite{Abgrall2017b,Bugg2014}. 

The low energy thresholds, excellent energy resolution, reduced electronic noise, and pulse shape characteristics of the P-type point contact detectors~\cite{luk89,Barbeau2007,Aguayo2011,Cooper2011} enable the sensitive double beta decay search. The nucleon decay analyses presented here include data taken from June 2015 until April 2018. Excluding calibration, commissioning data and data taken during intense mechanical work, the analyzed data includes 26.0 kg yr of enriched exposure and 9.45 kg yr of natural exposure~\cite{Alvis2018a}. The data are divided into  data sets referred to as DS0 through DS6 and a detailed description of each set is given in Ref.~\cite{Aalseth2018}. All the analyses described here were developed on the data sets published in Ref.~\cite{Aalseth2018} (approximately 1/3 of the total) and then executed on the full data sets after unblinding. The data blinding scheme parses the data into open (25\% of run time) and blind (75\%) partitions~\cite{Alvis2018a}.

The \DEM\ records every pulse with two digitizer channels with different amplifications to permit studies of the energy spectrum from below 1 keV to above 10~MeV. This work analyzes the spectrum from 100 keV to saturation (about 11 MeV). Energy deposits above saturation are recorded within an overflow channel and identified with a dedicated tag.

\section{Tri-Baryon Decay in Ge Isotopes}

Due to the enrichment of the Ge in the \DEM, the isotope \nuc{76}{Ge} has the largest exposure and dominates the sensitivity to $\slashed{B}$. Therefore we describe the analysis of the tri-proton decay channel of \nuc{76}{Ge} in some detail here as an example. All searched-for signatures are summarized in Table~\ref{tab:signatures}. We report results for decays of all Ge isotopes present in the \DEM, \nuc{70,72,73,74,76}{Ge}. 

\begin{table*}[h]
\caption{A summary of the signatures of each decay channel for which the \MJ\ \DEM\ has sensitivity, specifying the energy and timing requirements for the successive decays. The invisible decay mode signatures are composed of two successive decays and hence have two energy constraints and one time constraint. The decay-mode specific signatures include an initial saturated event (not listed here), followed by one or more decays at the energies listed below. N.A. is shorthand for not applicable.}
\begin{center}
\begin{tabular}{|l|c|c|c|c|}
\hline\hline
\multicolumn{5}{|c|}{Invisible Decay Modes}	\\
\hline
Decay Mode												&	$\tau_1$ 		&			$E_1$ 				&$\tau_2$		& $E_2$				\\
\hline
 \nuc{76}{Ge}(ppp) $\rightarrow$ \nuc{73}{Cu}	$\rightarrow$ \nuc{73}{Zn}		&	N.A.				&$(2.0,6.6)$ MeV		& $\Delta T < 117$ s		& $(2.0,4.3)$ MeV			\\
 \nuc{76}{Ge}(pp) $\rightarrow$ \nuc{74}{Zn} $\rightarrow$ \nuc{74}{Ga}		&	N.A.				&$(2.0,2.3)$ MeV		& $\Delta T < 40$ m		& $(2.0,5.4)$ MeV					\\
\nuc{74}{Ge}(ppp) $\rightarrow$ \nuc{71}{Cu} $\rightarrow$ \nuc{71}{Zn}		&	N.A.				&$(2.0,4.6)$ MeV	 	& $\Delta T < 12.5$ m		& $(2.0,2.8)$ MeV				\\
\hline
\multicolumn{5}{|c|}{Decay-Specific Modes}	\\
\hline
Decay Mode											&	$\tau_1$ 				&		$E_1$ 			&$\tau_2$				& $E_2$				\\
\hline
\nuc{76}{Ge}(ppp)$\rightarrow$ \nuc{73}{Cu} $ e^+\pi^+\pi^+$		&	$\Delta T < 21$ s		&	$(0.1,6.6)$ MeV		&	$\Delta T < 117$ s	& $(0.1,4.3)$ MeV				\\
\nuc{76}{Ge}(ppn) $\rightarrow$ \nuc{73}{Zn} $ e^+\pi^+$			&	$\Delta T < 117$ s		&	$(0.1,4.3)$ MeV		& N.A.				&	N.A.							\\
\nuc{76}{Ge}(pp) $\rightarrow$ \nuc{74}{Zn} $ \pi^+\pi^+$			&	$\Delta T < 4.5$ m		&	$(0.1,2.3)$ MeV		&	$\Delta T < 40$ m	& $(0.1,5.4)$ MeV				\\
\nuc{76}{Ge}(pn) $\rightarrow$ \nuc{74}{Ga} $ \pi^0\pi^+$			&	$\Delta T < 40$ m		&	$(0.1,5.4)$ MeV		& N.A.				&	N.A.							\\
\hline
\nuc{74}{Ge}(ppp) $\rightarrow$ \nuc{71}{Cu} $ e^+\pi^+\pi^+$		&	$\Delta T < 100$ s		&	$(0.1,4.6)$ MeV		&	$\Delta T < 12.5$ m	& $(0.1,2.8)$ MeV			\\
\nuc{74}{Ge}(ppn) $\rightarrow$ \nuc{71}{Zn} $ e^+\pi^+$			&	$\Delta T < 12.5$ m		&	$(0.1,2.8)$ MeV		& N.A.				&	N.A.							\\
\hline
\nuc{73}{Ge}(ppp) $\rightarrow$\nuc{70}{Cu} $ e^+\pi^+\pi^+$		&	$\Delta T < 25$ s		&	$(0.1,6.6)$ MeV		& N.A.				&	N.A.							\\
\nuc{73}{Ge}(pnn) $\rightarrow$\nuc{70}{Ga} $ e^+\pi^0$		&	$\Delta T < 105$ m		&	$(0.1,1.7)$ MeV		& N.A.				&	N.A.							\\
\nuc{73}{Ge}(pp) $\rightarrow$\nuc{71}{Zn} $ \pi^+\pi^+$			&	$\Delta T < 12.5$ m		&	$(0.1,2.8)$ MeV		& N.A.				&	N.A.							\\
\hline
\nuc{72}{Ge}(ppp) $\rightarrow$\nuc{69}{Cu} $ e^+\pi^+\pi^+$		&	$\Delta T < 15$ m		&	$(0.1,2.7)$ MeV		& N.A.				&	N.A.							\\
\nuc{72}{Ge}(pn) $\rightarrow$\nuc{70}{Ga} $ \pi^0\pi^+$			&	$\Delta T < 105$ m		&	$(0.1,1.7)$ MeV		& N.A.				&	N.A.							\\
\hline
\nuc{70}{Ge}(nnn) $\rightarrow$\nuc{67}{Ge} $\overline{\nu}\pi^0$	&	$\Delta T < 95$ m		&	$(0.1,4.4)$ MeV		& N.A.				&	N.A.							\\
\hline\hline
\end{tabular}
\end{center}
\label{tab:signatures}
\end{table*}%

The two analyses described here, invisible decay modes and decay-specific modes, are similar but have minor differences arising from the relative signature efficiency optimization. The signature for an invisible decay mode is the sequence of decays of the resulting unstable daughter, ignoring any potential signature from the initial disappearance of the nucleons. In the decay-specific mode searches, the decays of the unstable daughter nuclei are sought following an initial signature from the $\slashed{B}$ decay. For the \DEM\ the most sensitive channel, in both the decay-specific and invisible modes, is the tri-proton decay of \nuc{76}{Ge} to \nuc{73}{Cu}. The resulting \nuc{73}{Cu} isotope is $\beta$ unstable with a 4.2 s half-life and a Q-value of 6.6 MeV. Its daughter \nuc{73}{Zn} is also $\beta$ unstable with a 23.5 s half-life and a Q-value of 4.3 MeV. Since the count rate is very low in the \DEM\ above the two-neutrino double-beta decay endpoint (2 MeV), a signature of two $\beta$ decay candidates occurring within five half-lives (117 s) of one another, each above 2 MeV, has very little background. 

We chose a high-efficiency, five half-life time window between events to select candidate delayed coincidences. The average time between events with energy greater than 100 keV in a typical \DEM\ detector is $\approx$3 h and the decays of some long-lived isotopes were not considered due to potential accidental coincidence background. To keep the expected accidental background below 1 count with our time cut criterion, only isotopes with a half-life of $<$40 m were considered. This excluded consideration of the di-nucleon decays of \nuc{74}{Ge}, for example. In practice the longest coincidence window we considered was 105 m, corresponding to the 21 m half-life of \nuc{70}{Ga}.

\section{Invisible Decay Processes}
To select candidate events for invisible decays, we remove events in coincidence with the muon veto and those that fail the delayed-charge recovery (DCR) cut. The use of the DCR cut for this subset of the analysis reduces background due to alpha particles originating from near the detector surface. We do {\em not} reject multi-detector events or those waveforms symptomatic of multi-site events as some $\gamma$s might deposit energy in multiple locations. All these cuts are described in detail in Ref.~\cite{Aalseth2018} and references therein.  We then require energy and timing correlations between successive events within a lone detector to match a particular decay candidate. (See Table~\ref{tab:signatures}.) 

The total efficiency ($\epsilon_{tot}$) is equal to the product of all the efficiencies due to the time correlation cuts and the energy cuts. For the invisible decay modes, we study signatures with two beta decays.  The efficiency of the cut due to the decay of the second beta emitter is referred to as $\epsilon_{\tau2}$. (Note that $\epsilon_{\tau1}$ plays no role in the analysis of the invisible decay modes as there is no indicator for the creation of the first nucleus. This is in contrast to the decay specific modes discussed below.) 

\begin{figure}
\centering
\includegraphics[width=0.5\textwidth, height=0.2\textheight]{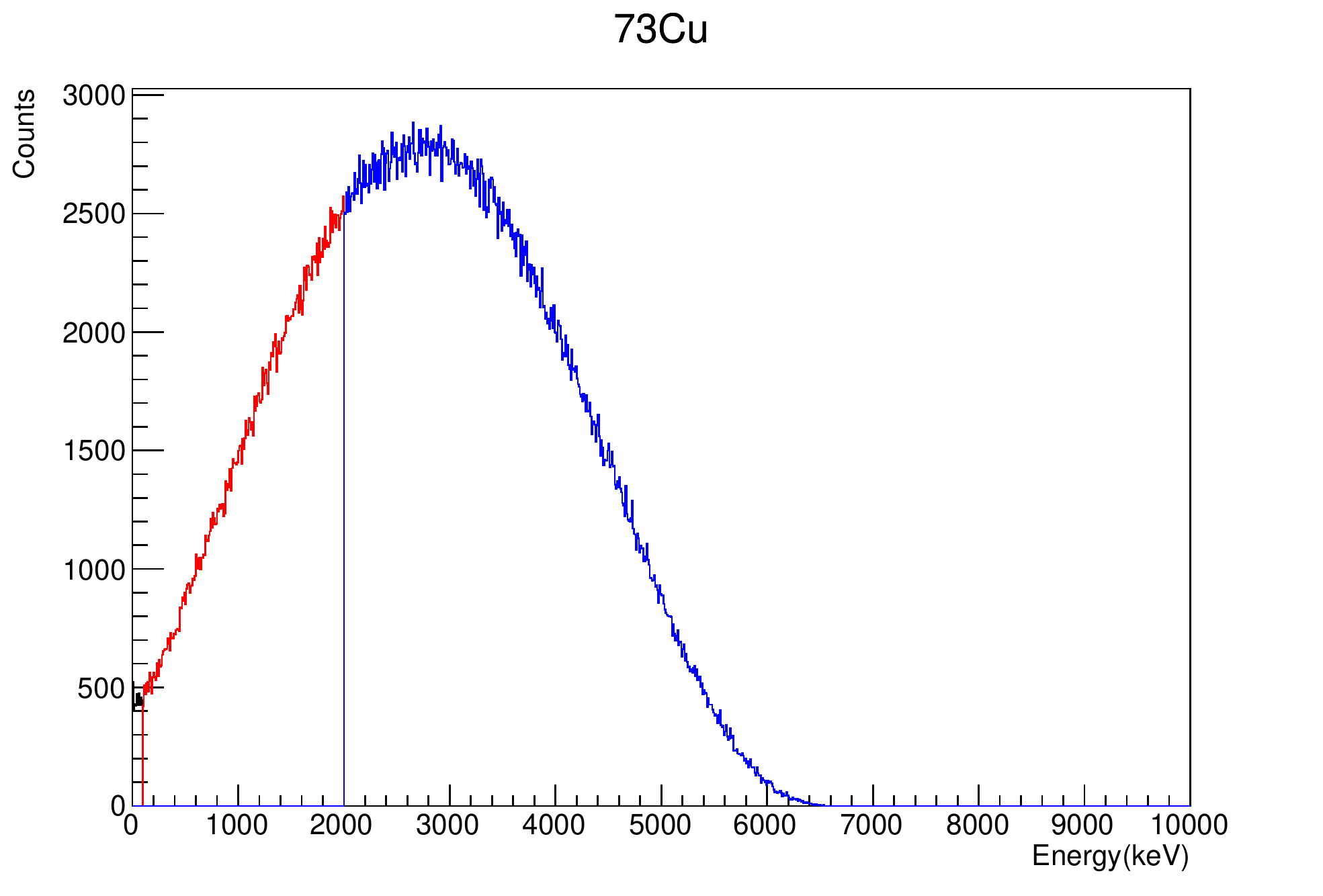}
\includegraphics[width=0.5\textwidth, height=0.2\textheight]{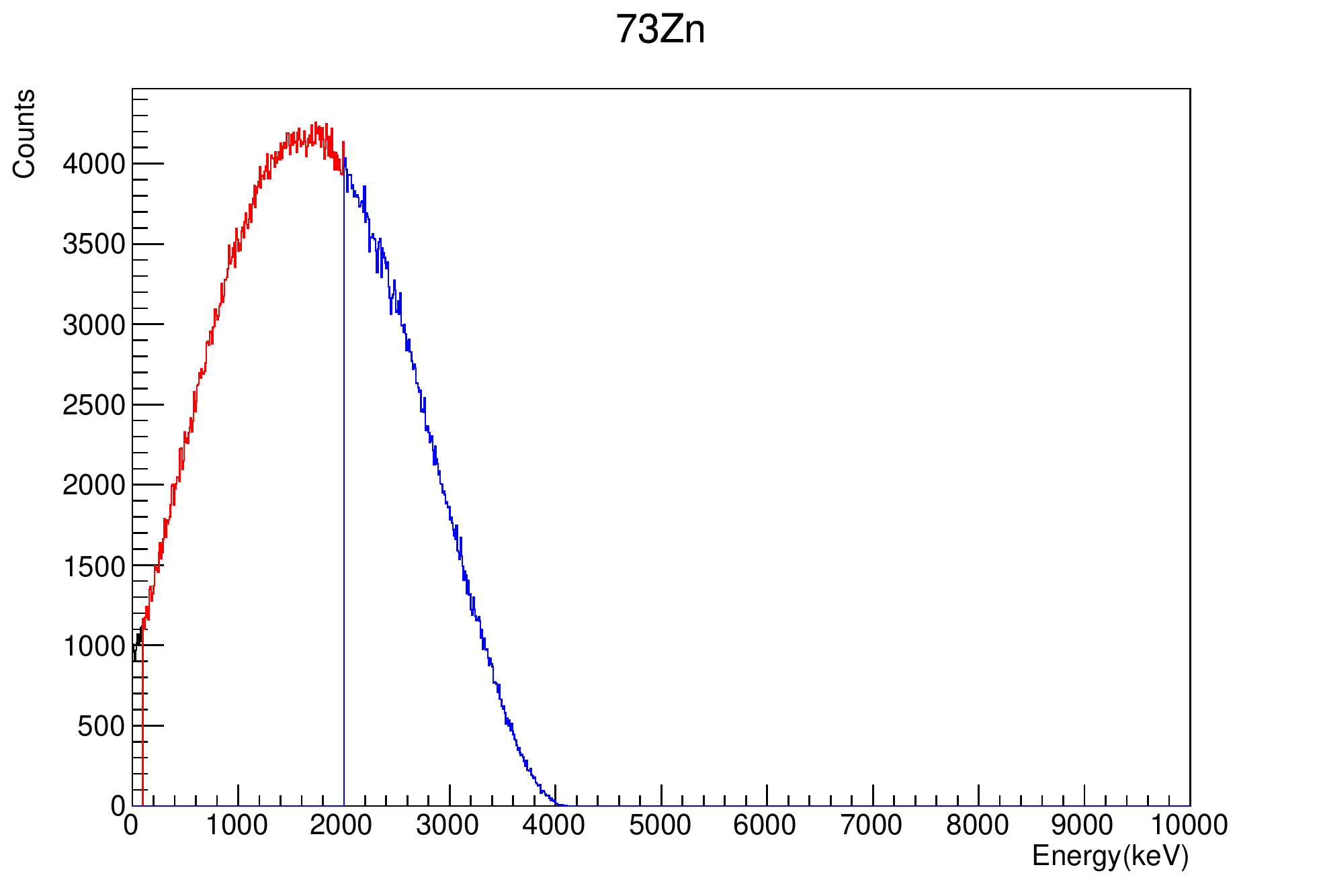}
\caption{The simulated energy deposit due to (Top)$^{73}$Cu and (Bottom)$^{73}$Zn. The fraction of the spectrum above 2 MeV for $^{73}$Cu decay is 70.7\% and for $^{73}$Zn decay is 37.5\%. The fraction of the spectrum above 100 keV for $^{73}$Cu decay is 99.6\% and for $^{73}$Zn decay is 99.0\%.}
\label{fig:energydeposit}
\end{figure}

For the invisible decay, a {\sc Geant}4-based\cite{Agostinelli2003250} Monte Carlo simulation framework ({\sc MaGe})~\cite{Chan:2008,Boswell2011}, was used to study the efficiency of the $\beta$/$\gamma$ decays depositing energy above the threshold. The decay manager within {\sc Geant}4 was used to simulate each isotope decay including branchings to excited states. For each isotope, we generated 1 million events in a detector and constrained the decay chain only to its daughter but no further.  Figure~\ref{fig:energydeposit} shows an example of the simulated energy spectrum of $^{73}$Cu and $^{73}$Zn decays in the detector. We calculated the efficiency ($\epsilon_{E_1}$) as the fraction of the events with energy larger than 2 MeV deposited. The {\sc MaGe} simulation framework has been vetted by comparison to \textsc{Majorana} $^{228}$Th calibration~\cite{Abgrall2017} and is found to describe the detector response very well. At energies of relevance here above 100 keV, the agreement is better than 2\%. It is even better if only one detector responds or when the energies are larger than 500 keV.

The time cut efficiency takes into account the boundaries of data acquisition periods. We define the efficiencies corresponding to the energy restrictions on the two $\beta$ decays as $\epsilon_{E1}$ and $\epsilon_{E2}$ corresponding to the first and second decay respectively. For the invisible decay modes, $\epsilon_{tot}=\epsilon_{E1}\epsilon_{\tau 2}\epsilon_{E2}\epsilon_{DCR}^2$, where $\epsilon_{DCR}$ represents a delayed charge recovery (DCR) waveform cut that rejects $\alpha$ induced signals~\cite{Gruszko2016}.

The half-life limit ($T_{1/2}$) is
\begin{equation}
T_{1/2}  > \frac{\ln(2) N T \epsilon_{tot}}{S},
\label{eq:halflife}
\end{equation}
where $N$ is the number of isotopic atoms within the detector active volume and $T$ is the live time in years. We found one such candidate for \nuc{76}{Ge} decay and used the Feldman-Cousins limit~\cite{Feldman1998} to set an upper limit on the number of events that could be assigned to the process of $S$=4.36 at the 90\% confidence-level half-life limit (Eqn.~\ref{eq:halflife}).  The efficiency for this signature ($\epsilon_{tot}$=0.257) includes factors due to the fraction of the beta decays with energy greater than 2 MeV, ($\epsilon_{E1}$=0.707, $\epsilon_{E2}$=0.375),  and the five half-life time restriction ($\epsilon_{\tau2}$=0.969) on the time difference between the two energy deposits, corresponding to the half-life $\tau_2$ in this case. In addition each of the two waveforms must survive the DCR cut. This efficiency ($\epsilon_{DCR} \sim 0.99$ for each waveform) varies from data set to data set but is near this nominal value. We account for the variation in the calculation of the product of efficiency and exposure.

\begin{figure}[!h]
 \centering
 \includegraphics[width=0.95\columnwidth, keepaspectratio=true]{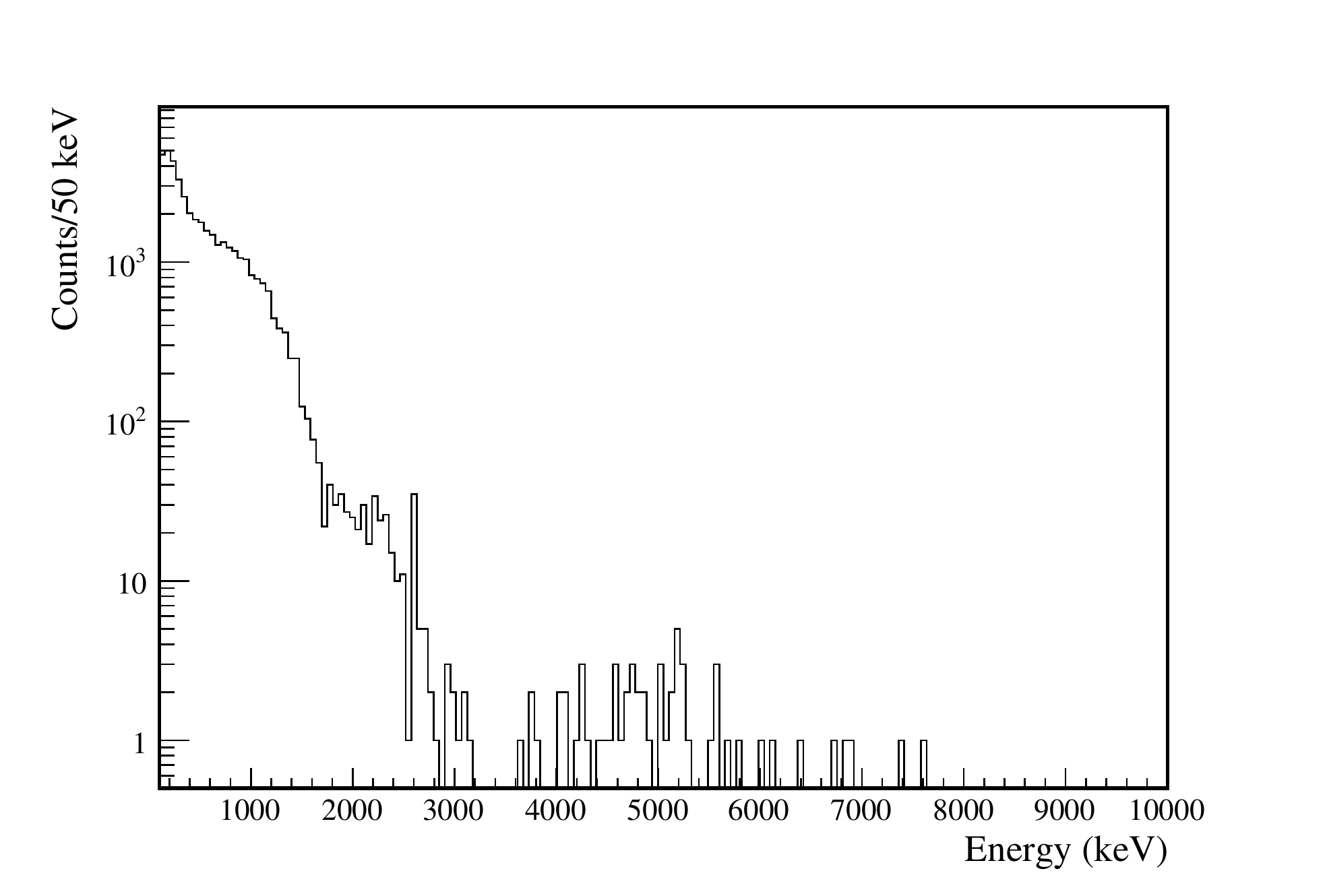}
 \includegraphics[width=0.95\columnwidth, keepaspectratio=true]{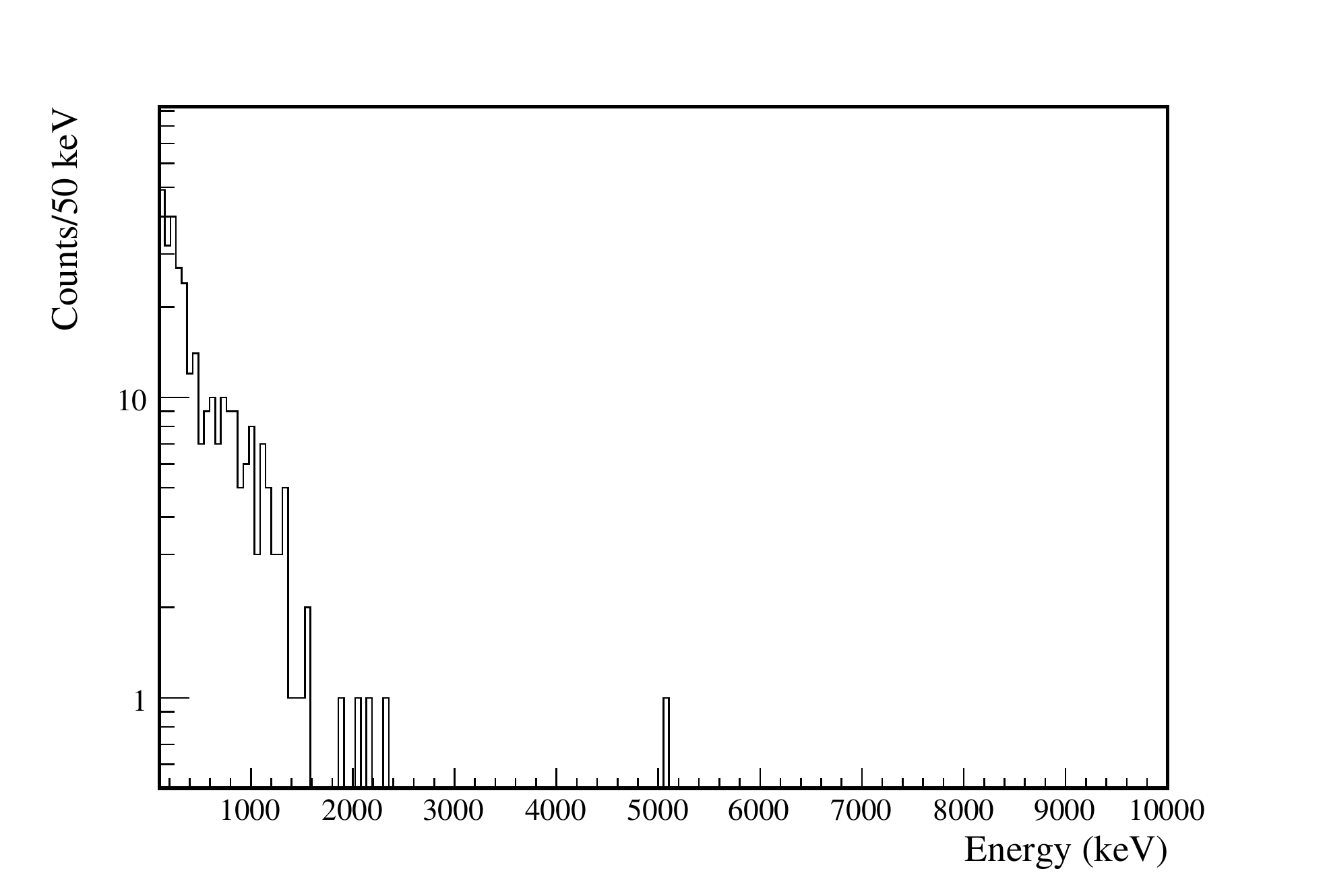}
 \caption{Top: The spectrum of all events surviving after a muon veto cut  and a DCR cut with energy greater than 100 keV that follow a previous event with energy greater than 100 keV in a given detector within a 40-minute delayed coincidence window. Bottom: The same as the top spectrum, except that the initial event is required to have at least 2 MeV, corresponding to one of the energy restrictions for candidates for invisible decay modes.  Of the 4 events above 2 MeV, only 2 (described in Table~\ref{tab:InvCandidates}) meet the combined requirements of energy and time to be candidates.  }
 \label{Fig:spectra}
\end{figure}

We perform a similar analysis for the invisible di-proton decay of \nuc{76}{Ge} and the tri-proton decay of \nuc{74}{Ge}. Table~\ref{tab:InvCandidates} lists the 2 events which can be considered candidates for any of these three invisible decay channels. The half-life limit results are given  in Table~\ref{tab:Limits}. Figure~\ref{Fig:spectra} shows the delayed coincidence spectra indicating the low background for these processes once the various cuts are implemented.

\begin{table}[!h]
\caption{The 2 candidate events for the invisible decays indicating processes to which they correspond. We assume each event is likely to be background for the indicated process when we calculate half-life limits. The \nuc{76}{Ge}(pp) and \nuc{76}{Ge}(ppp) processes each have 1 corresponding event. The \nuc{74}{Ge}(ppp) process has 2.}
\begin{center}
\begin{tabular}{|c|c|c|c|c|}
\hline\hline
Event	&		$E_1$ 		&	$E_2$ 	& $\tau_2$		 & Candidate 	 			\\
		&		(keV) 		&	 (keV) 	& 				 & 	Process(es) 			\\
\hline
1		&	4085	 			&	2164 	& $\Delta T = 12.9$ s			&	\nuc{76}{Ge}(ppp), \nuc{74}{Ge}(ppp)	\\
2		&	2092 			&	2353 	& $\Delta T = 2.7$ m			&	\nuc{76}{Ge}(pp), \nuc{74}{Ge}(ppp)	\\
\hline\hline
\end{tabular}
\end{center}
\label{tab:InvCandidates}
\end{table}%

\section{Decay Mode Specific Processes}
For decay modes specific to one of the processes in Eqns.~\ref{eqn:ThreeParticle} and \ref{eqn:TwoParticle},  the signature benefits from the energy deposit of the initial decay process ($\epsilon_0$) and the time correlation with the following decay of the unstable nucleus ($\epsilon_{\tau1}$). The decays in Eqns.~\ref{eqn:ThreeParticle} and \ref{eqn:TwoParticle} also have significant nuclear recoil kinetic energy, up to many 10's of MeV. A threshold of 11 MeV, chosen to lie above most of our events and near or at the digitizer saturation level, was applied to select these events. Even though edge effects can sometimes result in a modest lepton or pion energy deposit, the probability that the initial decay deposits more than 11 MeV is over 95\% for all decay channels. 

\begin{figure}
\centering
\includegraphics[width=0.5\textwidth, height=0.2\textheight]{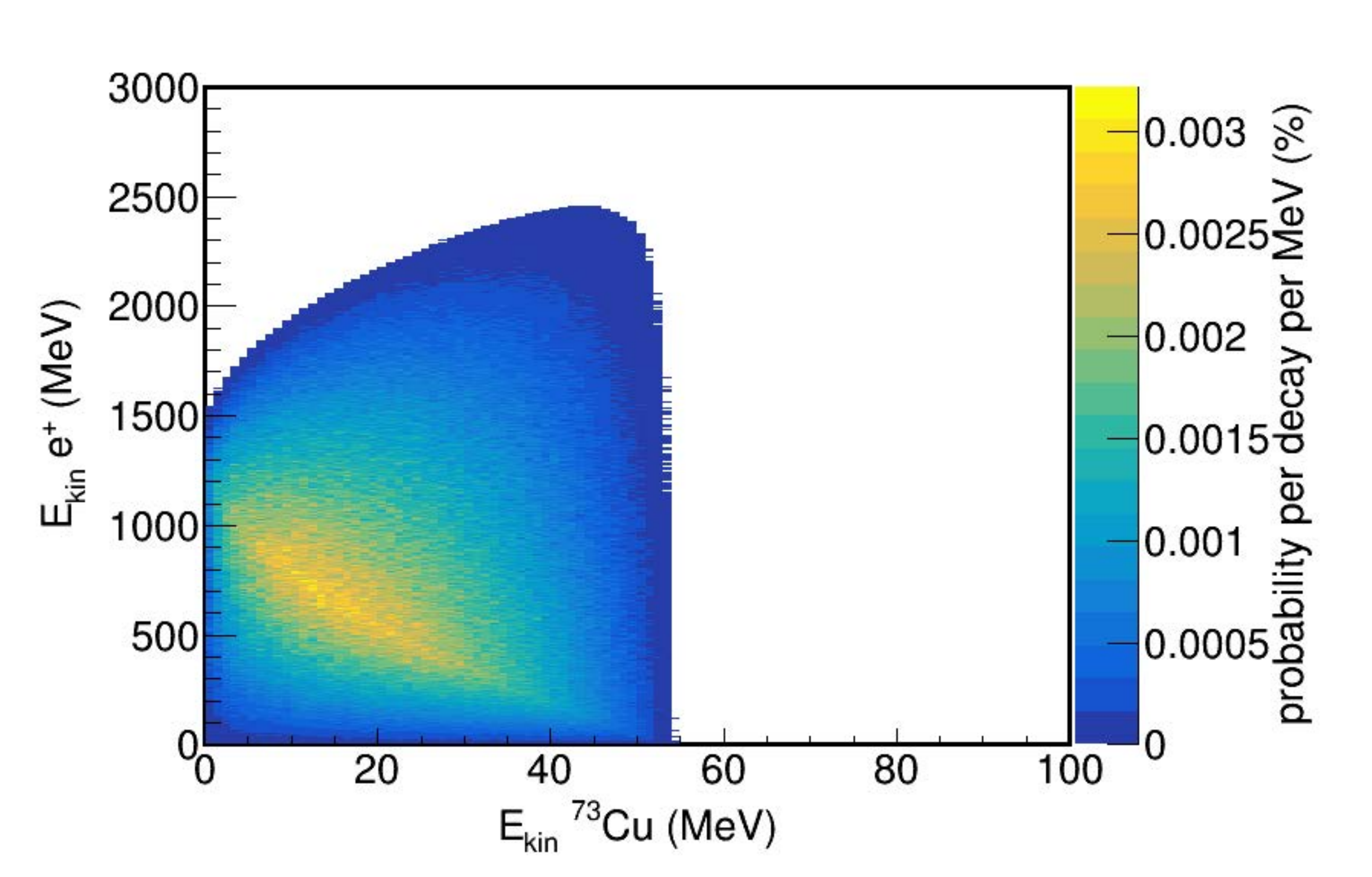}
\includegraphics[width=0.5\textwidth, height=0.2\textheight]{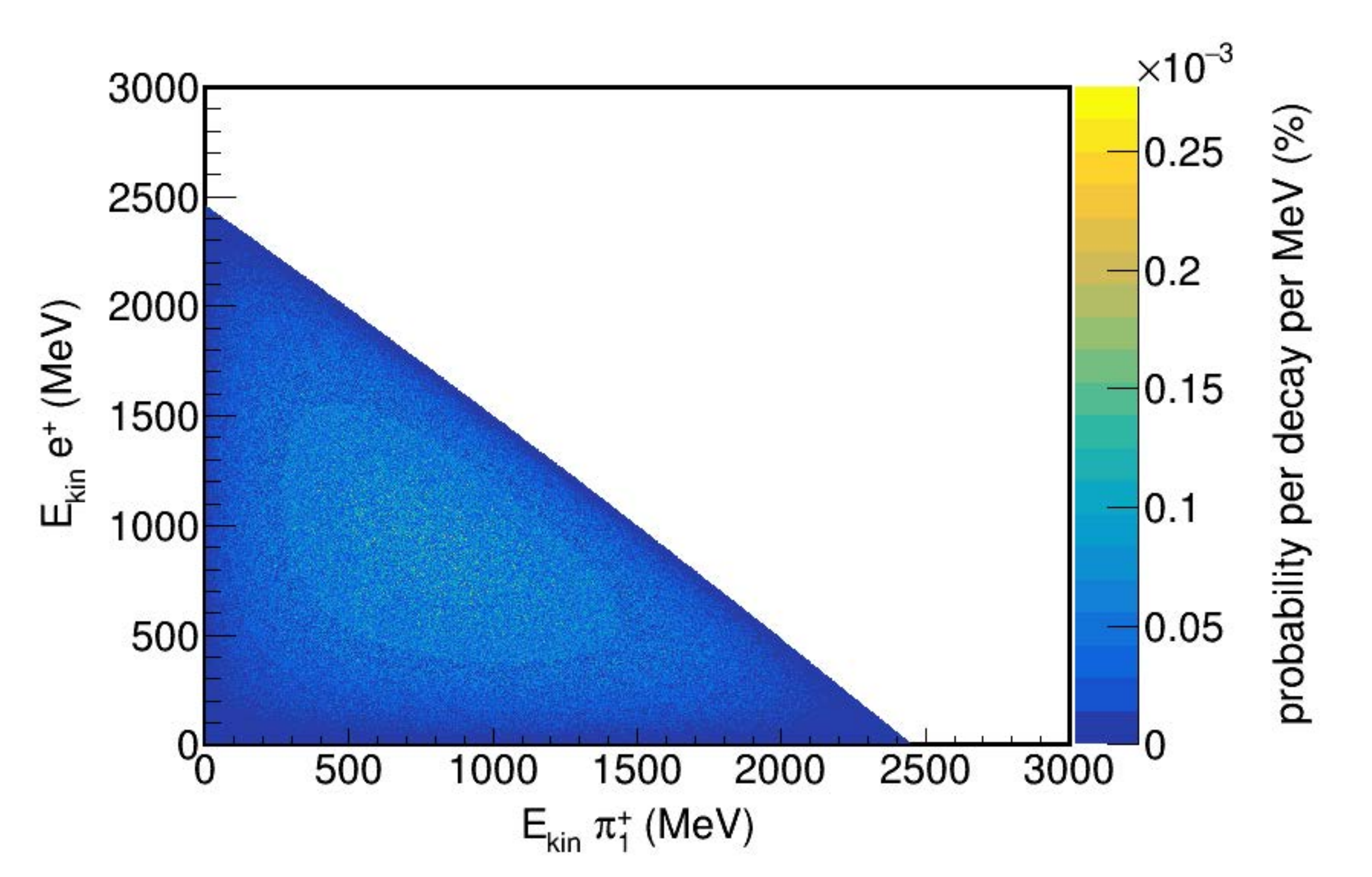}
\caption{The phase space distribution between particles in $^{76}\text{Ge} \rightarrow ^{73}\text{Cu}~e^{+}~\pi^{+}~ \pi^{+}$.}
\label{fig:phasespace_ppp}
\end{figure}

We used {\sc MaGe} to simulate these decay-mode-specific efficiencies also including all participating particles in Table~\ref{tab:Limits}. The emitted particles deposit a great deal of energy for the considered decays. The phase space distribution of the n-body decay was calculated using the GENBOD function~\cite{James:275743} in the TGenPhaseSpace class of ROOT~\cite{Brun:1997}. As an example,  Fig.~\ref{fig:phasespace_ppp} shows the phase space distribution for $^{76}\text{Ge}\rightarrow ^{73}\text{Cu} ~e^{+} ~\pi^{+} ~\pi^{+}$.  The efficiency was estimated as the fraction of the 10000 events with an energy larger than 11 MeV deposited within the detector. The nuclear recoil energy included a correction for  quenching using the Lindhard equation~\cite{Lindhard:1963}, but at these high energies, the shift in efficiency was less than the statistical uncertainty of the simulation, implying that quenching is not an important effect. Almost all events will have a large probability of saturating the detectors as shown in Table~\ref{tab:Limits}.  Due to this additional saturated event tag, the 2-MeV threshold constraint used for the invisible decay search can be relaxed.  The energy threshold for the decay-specific modes is 100 keV, resulting in a significantly higher efficiency. 

Therefore, there is a high probability that the event will be very distinctive. Although some saturated events arise from electrical breakdown and not physical processes, the associated waveforms are distinct from a saturated physics events and the two populations can be easily discerned by pulse shape analysis. In particular the onset of the waveform of a physics event is gradual, whereas for a breakdown it is a sharp upturn. Cosmic rays are also a source of saturated waveforms, but the veto system tags them efficiently.

For the decay-specific modes, we remove non-physical waveforms but do not apply the DCR cut. The DCR cut is unnecessary because the saturated event trigger rate is very low, significantly reducing the background. For the decay-specific modes analyses, we also require full operation of the cosmic ray veto system as candidates will have a large energy deposit that is not muon induced. In DS0, the veto system was not fully implemented and we exclude that data from this analysis. This loss of exposure is accounted for in Table~\ref{tab:Limits}.  We then require energy and timing correlations between successive events, which differ from similar requirements for the invisible modes. 

The total efficiency ($\epsilon_{tot}$) is equal to the product of all the efficiencies due to the time correlation cuts, the energy cuts, and the efficiency for the detection of the initial decay ($\epsilon_0$). For the decay-specific modes, $\epsilon_{tot}=\epsilon_0\epsilon_{\tau1}\epsilon_{\tau2}\epsilon_{E1}\epsilon_{E2}$. Some processes we considered here only have one $\beta$ decay; in these cases $\epsilon_{\tau 2}$ and $\epsilon_{E2}$ are not applicable.

\begin{table*}[!h]
\caption{Efficiencies, exposures, signal upper limits and half-life limits for the modes of nucleon decay for the Ge isotopes for which the \DEM\ has an interesting sensitivity. The signal upper limit ($S$) is the Feldman-Cousins 90\% upper limit ($S$) given a number of observed candidates. N.A. is shorthand for not applicable.}
\begin{center}
\begin{tabular}{|l|c|c|c|c|c|c|c|c|c|c|}
\hline\hline
Decay Mode								&		$\epsilon_0$		&	$\epsilon_{\tau1}$	&$\epsilon_{E1}$	&$\epsilon_{\tau2}$		&$\epsilon_{E2}$	&$\epsilon_{tot}$	&NT$\epsilon_{tot}$ 		& Candidates	& $S$ 		& $T_{1/2}$ 	\\
										&						&					&					&				&				&				&(10$^{24}$ atom yr)	&			&(counts) 		&	($10^{24}$ yr)	\\
\hline
\multicolumn{10}{|c|}{Invisible Decay Modes}	\\
\hline
 \nuc{76}{Ge}(ppp) $\rightarrow$ \nuc{73}{Cu}&			N.A.					&	N.A.			&	0.707			&	0.969			&	0.375		&	0.26				&47.1			& 1	&	4.36		&\phantom07.5\\
 \nuc{76}{Ge}(pp) $\rightarrow$ \nuc{74}{Zn}	&		N.A.					&	N.A.			&	0.004			&	0.969			&	0.367		&	0.002			&\phantom0\phantom00.28& 1		&4.36 	&\phantom0\phantom00.05\\
\nuc{74}{Ge}(ppp) $\rightarrow$ \nuc{71}{Cu}&	 		N.A.					&	N.A.			& 	0.411			&	0.969			&	0.073		&	0.03				&\phantom01.5		&2	&	5.91			& \phantom0\phantom00.18\\
\hline
\multicolumn{10}{|c|}{Decay-Specific Modes}	\\
\hline
\nuc{76}{Ge}(ppp)$\rightarrow$ \nuc{73}{Cu} $ e^+\pi^+\pi^+$&	0.998			&	0.969			&	0.996		&	0.969			&	0.990		&	0.923		&165.				&0&	2.44		&	47.0\\
\nuc{76}{Ge}(ppn) $\rightarrow$ \nuc{73}{Zn} $ e^+\pi^+$	&		0.999		&	0.969			&	0.990		&	N.A.				&	N.A.			&	0.958		&172.				&0&	2.44		&	48.7	\\
\nuc{76}{Ge}(pp) $\rightarrow$ \nuc{74}{Zn} $ \pi^+\pi^+$	& 		0.994		&	0.968			&	0.972		&	0.964			&	0.991		&	0.893		&160.				&0&	2.44		& 45.5	\\
\nuc{76}{Ge}(pn) $\rightarrow$ \nuc{74}{Ga} $ \pi^0\pi^+$	&		0.979		&	0.964			&	0.991		&	N.A.				&	N.A.			&	0.935		&168.				&0&	2.44		& 47.6	\\
\hline
\nuc{74}{Ge}(ppp) $\rightarrow$ \nuc{71}{Cu} $ e^+\pi^+\pi^+$&	0.998		&	0.969			&	0.993		&	0.969			&	0.982		&	0.912		& 46.6				&0&	2.44		&13.2	\\
\nuc{74}{Ge}(ppn) $\rightarrow$ \nuc{71}{Zn} $ e^+\pi^+$	&		0.999		&	0.967			&	0.982		&	N.A.				&	N.A.			&	0.949		&48.5				&0&	2.44		&13.8\\
\hline
\nuc{73}{Ge}(ppp) $\rightarrow$\nuc{70}{Cu} $ e^+\pi^+\pi^+$&		0.998		&	0.968			&	0.996		&	N.A.				&	N.A.			&	0.963		&\phantom05.3			&0&	2.44		&\phantom01.5\\
\nuc{73}{Ge}(pnn) $\rightarrow$\nuc{70}{Ga} $ e^+\pi^0$&		0.999		&	0.958			&	0.867		&	N.A.				&	N.A.			&	0.830		&\phantom04.6			&1&	4.36		&\phantom00.7	\\
\nuc{73}{Ge}(pp) $\rightarrow$\nuc{71}{Zn} $ \pi^+\pi^+$	&		0.994		&	0.967			&	0.982		&	N.A.				&	N.A.			&	0.944		&\phantom05.2			&0&	2.44		&\phantom01.5	\\
\hline
\nuc{72}{Ge}(ppp) $\rightarrow$\nuc{69}{Cu} $ e^+\pi^+\pi^+$&		0.998		&	0.967			&	0.973		&	N.A.				&	N.A.			&	0.940		&18.4				&0&	2.44		&\phantom05.2	\\
\nuc{72}{Ge}(pn) $\rightarrow$\nuc{70}{Ga} $ \pi^0\pi^+$	&		0.979		&	0.958			&	0.867		&	N.A.				&	N.A.			&	0.813		&16.0				&1&	4.36		&\phantom02.5	\\
\hline
\nuc{70}{Ge}(nnn) $\rightarrow$\nuc{67}{Ge} $\overline{\nu}\pi^0$&	0.952		&	0.959			&	0.972		&	N.A.				&	N.A.			&	0.887		&11.9				&1&	4.36		&\phantom01.9\\
\hline\hline
\end{tabular}
\end{center}
\label{tab:Limits}
\end{table*}%

There is only one event with energy $>11$ MeV that meets the criteria to be a candidate. This event has a secondary energy deposition of 152 keV that follows the saturated event by 75.7 m. That event candidate matches the signature for three processes, \nuc{73}{Ge}(pnn), \nuc{72}{Ge}(pn), and \nuc{70}{Ge}(nnn), providing background for each.  The other searched-for channels have zero candidates. The $T_{1/2}$ limits for 12 different decay-specific modes are listed in Table~\ref{tab:Limits}.

\section{Discussion}
The systematic uncertainties include the exposure uncertainty (2\%), uncertainty in the non-physical event removal (0.1\%), uncertainty in the delayed charge recovery cut energy dependence (1\%),  uncertainty due to how well the simulations model the detector (2\%) and the statistical uncertainty of the simulated efficiencies ($<$1\%). All of these are very small compared to the statistical uncertainty of $S$, and we ignore their contribution to the half-life limits. The simulations were for specific modes of decay and hence have that model dependency as an uncertainty, however, we quote limits for the specific modes simulated. 
We find no evidence for $\slashed{B}$ and the best limits for the various decay-specific modes are mid 10$^{25}$ yr range.  The best limit for an invisible decay is for \nuc{76}{Ge}(ppp) $\rightarrow$ \nuc{73}{Cu} with a half-life $>7.5\times10^{24}$ yr.

For the di-nucleon modes, the Fr\'{e}jus~\cite{Berger1991}, KamLAND~\cite{Araki2006} and Super-Kamiokande~\cite{Litos2014,Takhistov2015,Gustafson2015} experiments have limits exceeding 10$^{30}$ yr, reaching out to $4\times10^{32}$ yr. Neutron-antineutron oscillations are also a $\Delta B$=2 test of $\slashed{B}$. SNO~\cite{Aharmim2017} reported a half-life limit for \nuc{2}{H} of $1.48\times10^{31}$ yr and Super-Kamiokande~\cite{Abe2015} reported a half-life limit of $1.9\times10^{31}$ yr for \nuc{16}{O}. The \DEM\ limits for di-nulceon modes are much less restrictive than these previous efforts because of the lower exposure. We list the results, however, in case the nuclear dependence is of interest.

It should be noted that some previous results are quoted in terms of a baryon half-life by attempting to account for the number of baryon combinations within a nucleus. Others quote a nuclear half-life. We chose the latter approach as the experimental result has less dependence on the model and interpretation. Furthermore, our quoted limits for each decay channel assume it is the dominant decay branch. This results in a conservative upper limit on the half-life for the considered channel. For example, \nuc{73}{Cu} could be populated by two-proton decay of \nuc{76}{Ge} to unbound states in \nuc{74}{Zn}, which in turn emits a proton. This process would compete with the tri-proton decay of \nuc{76}{Ge}. We neglect such side channels and quote the conservative lower value for the limit. It is also possible that the decay would result in excited states in \nuc{73}{Cu}. In this case the relaxation of this state would either be in coincidence with the initial decay products or would simply be a precursor event to our search. In neither case would that alter our search algorithm or efficiencies.

The best previous limits on 3n decays ($1.8\times10^{23}$ yr)~\cite{Hazama1994} come from a study in iodine, which also reported results for 4n decay ($1.4\times10^{23}$ yr). This paper took account of the number of baryon combinations within the same shell orbit. 

The \MJ\ \DEM\ provides an improved limit for 3p invisible decay. The previous best limits on tri-nucleon decay come from EXO-200~\cite{Albert2018} based on 223 kg yr of exposure. For the decay of \nuc{136}{Xe}(ppp) $\rightarrow$ \nuc{133}{Sb}, the limit is 3.3$\times10^{23}$ yr. For \nuc{136}{Xe}(ppn) $\rightarrow$ \nuc{133}{Te}, the limit is 1.9$\times10^{23}$ yr. The energy and time-coincidence cuts permit an event-by-event analysis in the \DEM, greatly reducing the background while maintaining a substantial efficiency. This results in an improved sensitivity over a spectral component fit approach.

\section*{acknowledgments}
We thank Michael Graesser for discussions on baryon decay.

This material is based upon work supported by the U.S.~Department of Energy, Office of Science, Office of Nuclear Physics under Award Numbers DE-AC02-05CH11231,  DE-AC05-00OR22725, DE-AC05-76RL0130, DE-AC52-06NA25396, DE-FG02-97ER41020, DE-FG02-97ER41033, DE-FG02-97ER41041, DE-SC0010254, DE-SC0012612, DE-SC0014445, and DE-SC0018060. We acknowledge support from the Particle Astrophysics Program and Nuclear Physics Program of the National Science Foundation through grant numbers MRI-0923142, PHY-1003399, PHY-1102292, PHY-1206314, PHY-1614611, PHY-1812356, and  PHY-1812409. We gratefully acknowledge the support of the U.S.~Department of Energy through the LANL/LDRD Program and through the PNNL/LDRD Program for this work. We acknowledge support from the Russian Foundation for Basic Research, grant No.~15-02-02919. We acknowledge the support of the Natural Sciences and Engineering Research Council of Canada, funding reference number SAPIN-2017-00023, and from the Canada Foundation for Innovation John R.~Evans Leaders Fund.  This research used resources provided by the Oak Ridge Leadership Computing Facility at Oak Ridge National Laboratory and by the National Energy Research Scientific Computing Center, a U.S.~Department of Energy Office of Science User Facility. We thank our hosts and colleagues at the Sanford Underground Research Facility for their support.

\bibliography{BDKpaper}

\end{document}